\documentclass[fleqn,12pt,twoside]{article}
\usepackage[headings]{espcrc1}
\usepackage{graphics,graphicx,amsmath,array,hhline,fancyhdr}
\usepackage[latin2]{inputenc}
\usepackage[magyar,english]{babel}

\def\qinv{q_\textrm{inv}}

\begin{document}
\title{Measurement and analysis of two- and three-particle correlations}
\author{M.~Csan\'ad\address{Department of Atomic Physics, ELTE Budapest, Pázmány P. s. 1/a, H-1117}
        for the PHENIX\footnote{For the full list of PHENIX authors and
acknowledgments, see appendix 'Collaborations' of this volume.} Collaboration}
\runauthor{M.~Csan\'ad for the PHENIX collaboration}
\maketitle

\begin{abstract}
Allowed regions of core fraction versus partial coherence are
obtained from analysis of both two- and three-pion Bose-Einstein
correlations measured at PHENIX in $\sqrt{s_\textrm{NN}}=200$ GeV
Au+Au collisions. Two-pion Bose-Einstein correlation functions for
different average transverse momenta are used to analyze chiral
symmetry restoration.

\end{abstract}

\section{Introduction}

Correlation functions are important to see the collective
properties of particles and space-time structure of the emitting
source, eg. the observed size of a system can be measured by two
particle Bose-Einstein correlations~\cite{HBT}. The $m_t$
dependent strength of two-pion correlations provides information
about U$_\textrm{A}(1)$ symmetry restoration in the
source~\cite{Vance:1998wd,Kapusta:1995ww,Huang:1995fc,Hatsuda:1994pi}.
From a consistent analysis of two- and three-particle correlations
one can establish an experimental measure of thermalization and
coherence in the source~\cite{Csorgo:1998tn}.

\section{Data Analysis}

We studied 70M $\sqrt{s_\textrm{NN}}=200$ GeV RHIC Au+Au
collisions recorded at PHENIX during Run 4. Charged tracks were
detected by the Drift Chamber and Pad Chambers. The Time Of Flight
detector and the Electromagnetic Calorimeters were used to
identify $\pi^+$ particles, and we had 200M $\pi^+$'s, 900M
$\pi^+$ pairs and more than 4G~$\pi^+$ triplets in this analysis.
We measured from this data sample the two-pion correlation
function $C_2$ as a function of $\qinv = \sqrt{(k_1-k_2)^2}$ and
the three-pion correlation function $C_3$ as a function of $\qinv$
of the three pairs in the triplet, $q_{12}, q_{23}, q_{31}$.

Our method of Coulomb-correction is based on the solution of the
two-body Coulomb-problem. A symmetrized asymptotically correct
three-body Coulomb wave-function is built up from the two-body
solution. The source function is then integrated with the absolute
square of this three-body Coulomb-wave function as the density
function, and the same is done for a symmetrized three-body
plain-wave function. The ratio of the two gives the
Coulomb-correction factor~\cite{Alt:2001dj}. For the final
Coulomb-correction we utilized a core/halo type of picture of the
source and iterated $\lambda$ and $R$ in a self-consistent manner
until convergence.

The fits were done using three different shapes, Gauss, Levy and
Edgeworth, described in ref.~\cite{Csorgo:1999sj}. We fitted only
those points with $\qinv>0.02$ GeV$/c$ due to a non-BEC and
non-Coulomb structure seen at low $\qinv$ (see the gray lines on
fig.~\ref{f:pifit}).

\section{Results}

\begin{figure}\begin{center}
  \includegraphics[width=2.75in]{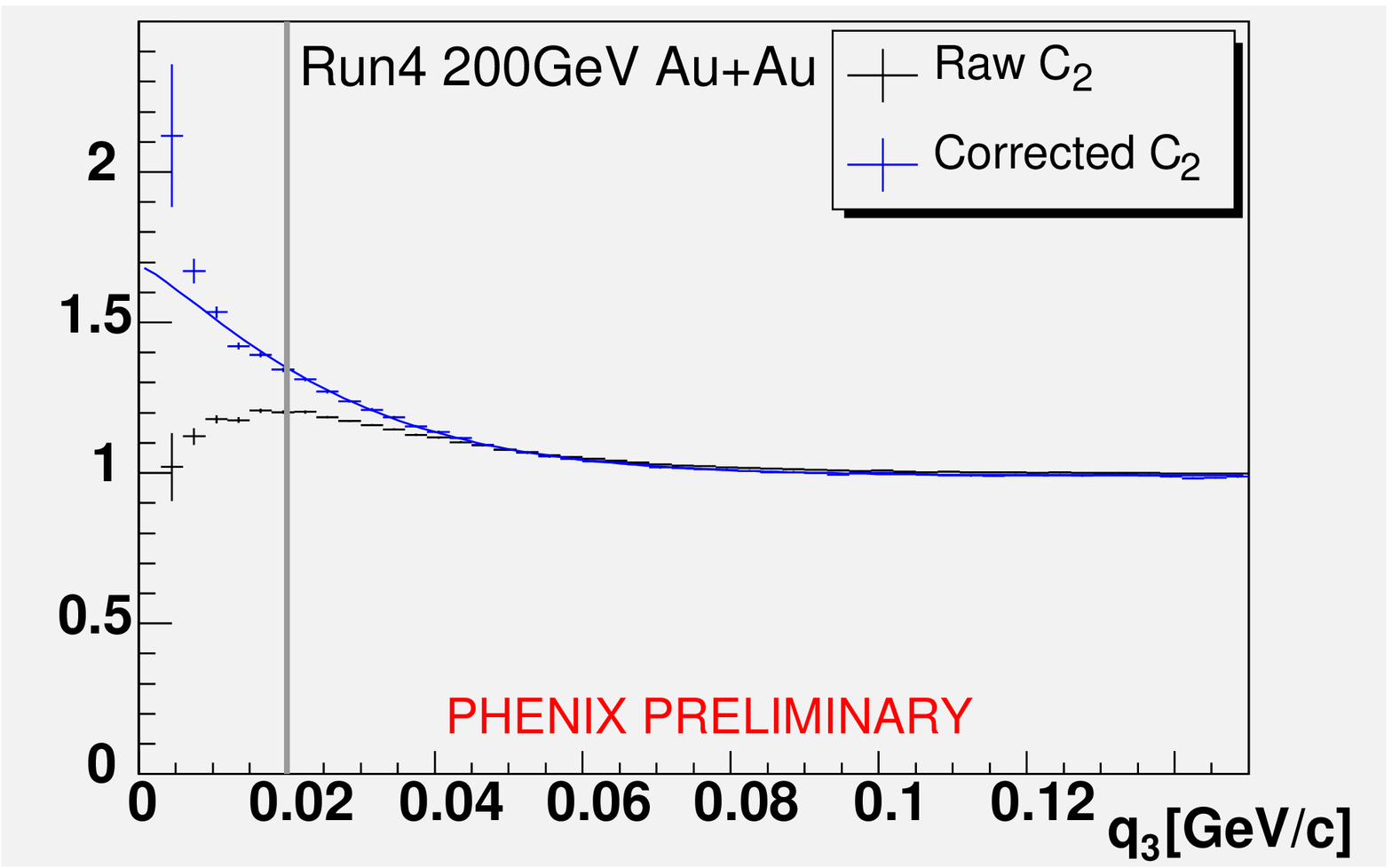} \includegraphics[width=2.75in]{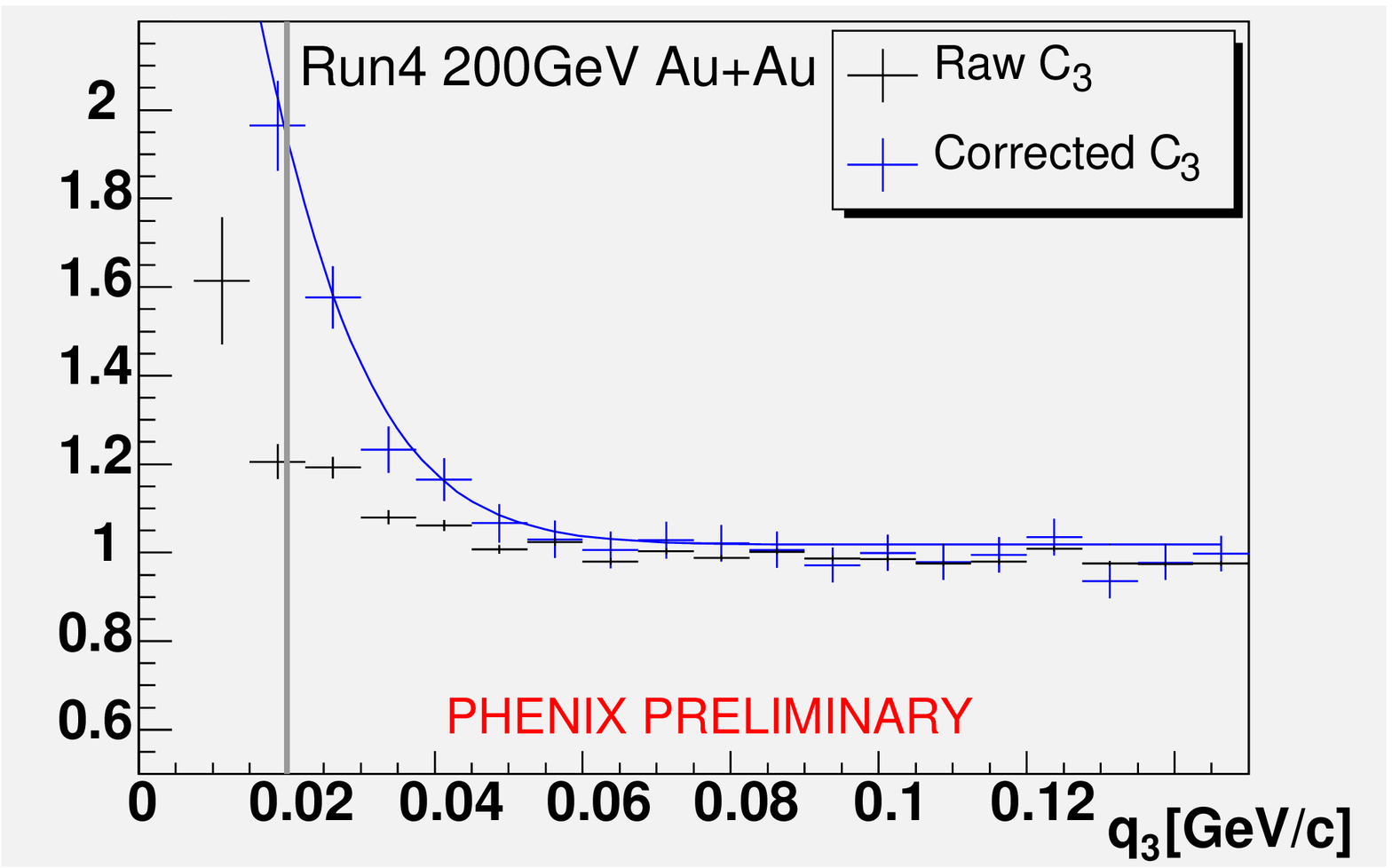}\\
  \caption{Raw- and Coulomb-corrected correlation functions for
  $0.2 < p_t < 2.0$ GeV$/c$ and $0-92\%$ centrality with Levy fits.
  Points with $\qinv<0.02$ GeV$/c$ were excluded from the fits.}
  \label{f:pifit}
\end{center}\end{figure}

First, we fitted two- and three-particle correlation functions for
$0.2 < p_t < 2.0$ GeV$/c$ and $0-92\%$ centrality. Here only the
Levy fits had acceptable confidence levels. These and the
Coulomb-corrected correlation functions are shown on
fig.~\ref{f:pifit}. The three-dimensional $C_3$ was projected onto
the $q_{12}=q_{23}=q_{31}$ line.

Let us consider the ratios $f_c=N_c(p)/N_1(p)$ (fractional core)
and $p_c=N_c^p(p)/N_c(p)$ (partial coherence), and take the
equations (see eqs.~57 and 58 in ref.~\cite{Csorgo:1999sj}):
\begin{eqnarray}
C_2(p_1 \simeq p_2)&=&1+f_c^2[1-p_c^2] \label{e:c2fcpc}\\
C_3(p_1 \simeq p_2 \simeq p_3)&=&1+3f_c^2[1-p_c^2]
+2f_c^3[(1-p_c)^3+3p_c(1-p_c)^2]\label{e:c3fcpc}
\end{eqnarray}

\begin{figure}\begin{center}
  \center
  \includegraphics[width=2.35in]{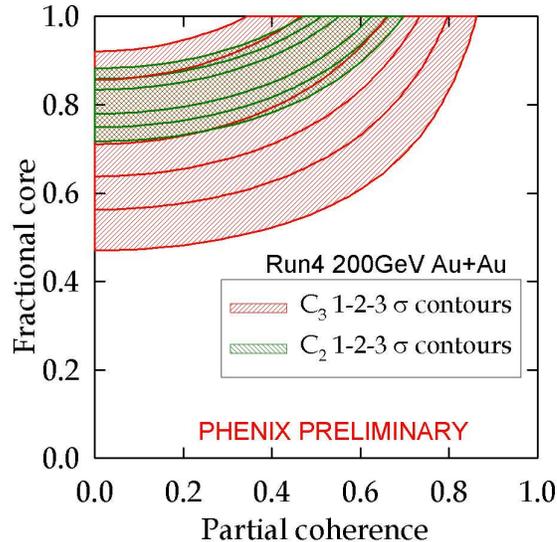}\\
  \caption{Allowed region of partial coherence versus fractional
  core extracted from Levy fits. Contours obtained from $C_3$ are
  much larger due to smaller statistics of $\pi^+$ triplets compared
  to $\pi^+$ pairs.}\label{f:fcpc}
\end{center}\end{figure}

From this we calculated the allowed $f_c$ vs $p_c$ region, using
the 3$\sigma$ contours from eqs.~\ref{e:c2fcpc}-\ref{e:c3fcpc},
see fig.~\ref{f:fcpc}. The size of the region obtained from $C_3$
is much larger due to the smaller number of triplets compared to
pairs.

Compared to former NA44 analysis (see
refs.~\cite{Csorgo:1998tn,Boggild:1999tu,Biyajima:2003ey}), the
resulting allowed region is at a little bit higher $f_c$ values,
because our $f_c$ was obtained from Levy fits and not from Gauss,
as confidence level of latter was unacceptable.

In the second part of the analysis, we have analyzed two-pion
correlation functions in 10 $p_t$ bins from 0.2 GeV$/c$ to 0.5
GeV$/c$ with $0-92\%$ centrality, using the invariant momentum
difference $q \equiv \qinv$ as a variable and the following shapes
(in $\hbar c = 1$ units):
\begin{eqnarray}
\textrm{Gauss: \;} C_2(q)&=&1+\lambda \exp(-\left|Rq\right|^{2})\\
\textrm{Levy: \;} C_2(q)&=&1+\lambda \exp(-\left|Rq\right|^{\alpha})\\
\textrm{Edgeworth: \;} C_2(q)&=&1+\lambda
\exp(-\left|Rq\right|^{2})
   (1+\frac{\kappa_3}{3!}H_3(\sqrt{2}\left|Rq\right|),
\end{eqnarray}
where $H_3$ is the third Hermite polynomial, and $\kappa_3$ it's
coefficient (the Edgeworth form is a model-independent expansion,
of which we keep only the first term, to measure deviation from
the Gauss form). See more details in ref.~\cite{Csorgo:2003uv}.

\begin{figure}\begin{center}
  \includegraphics[width=5.9in]{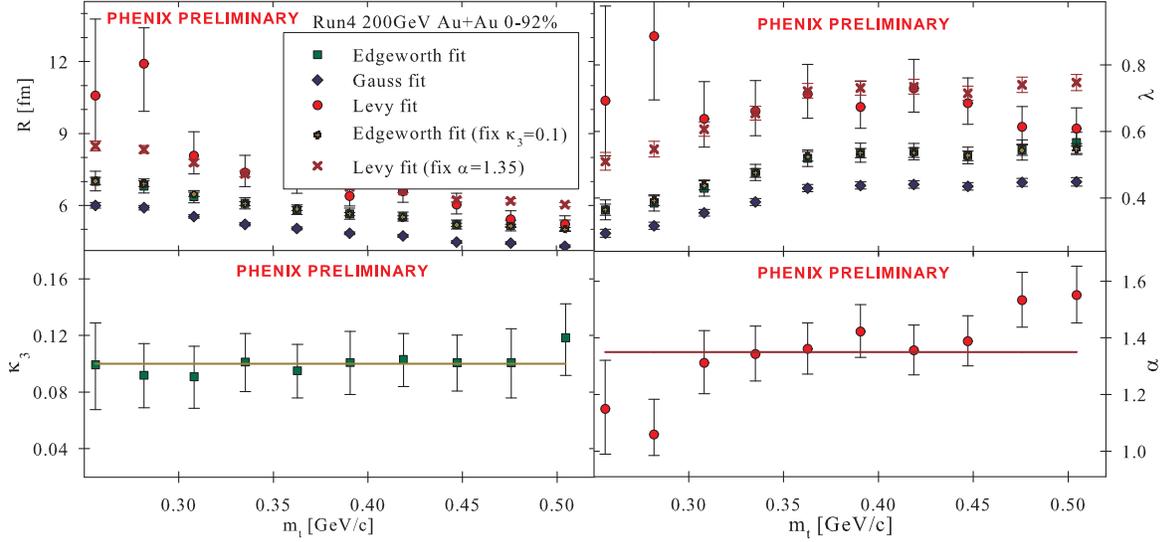}\\
  \caption{$\lambda(m_t)$, $R(m_t)$, $\alpha(m_t)$ and $\kappa_3(m_t)$ from Gauss, Levy and Edgeworth fits}
  \label{f:shapeall}
\end{center}\end{figure}

The parameters as a function of average transverse mass
($\lambda(m_t)$, $R(m_t)$, $\alpha(m_t)$ and $\kappa_3(m_t)$) for
Gauss, Levy and Edgeworth fits are shown on fig.~\ref{f:shapeall}.
The $\alpha$ parameter of the Levy distribution controls the
long-range source in a similar way as the $\lambda$ parameter, so
fitting both makes the fit underconstrained. Hence we did fits
also with fixed $\alpha$, and for comparison, also with fixed
$\kappa_3$. These latter fits show the same $\lambda(m_t)$
behavior as in the Gaussian case.

\begin{figure}[ht]
   \begin{center}
   \includegraphics[width=1.9in]{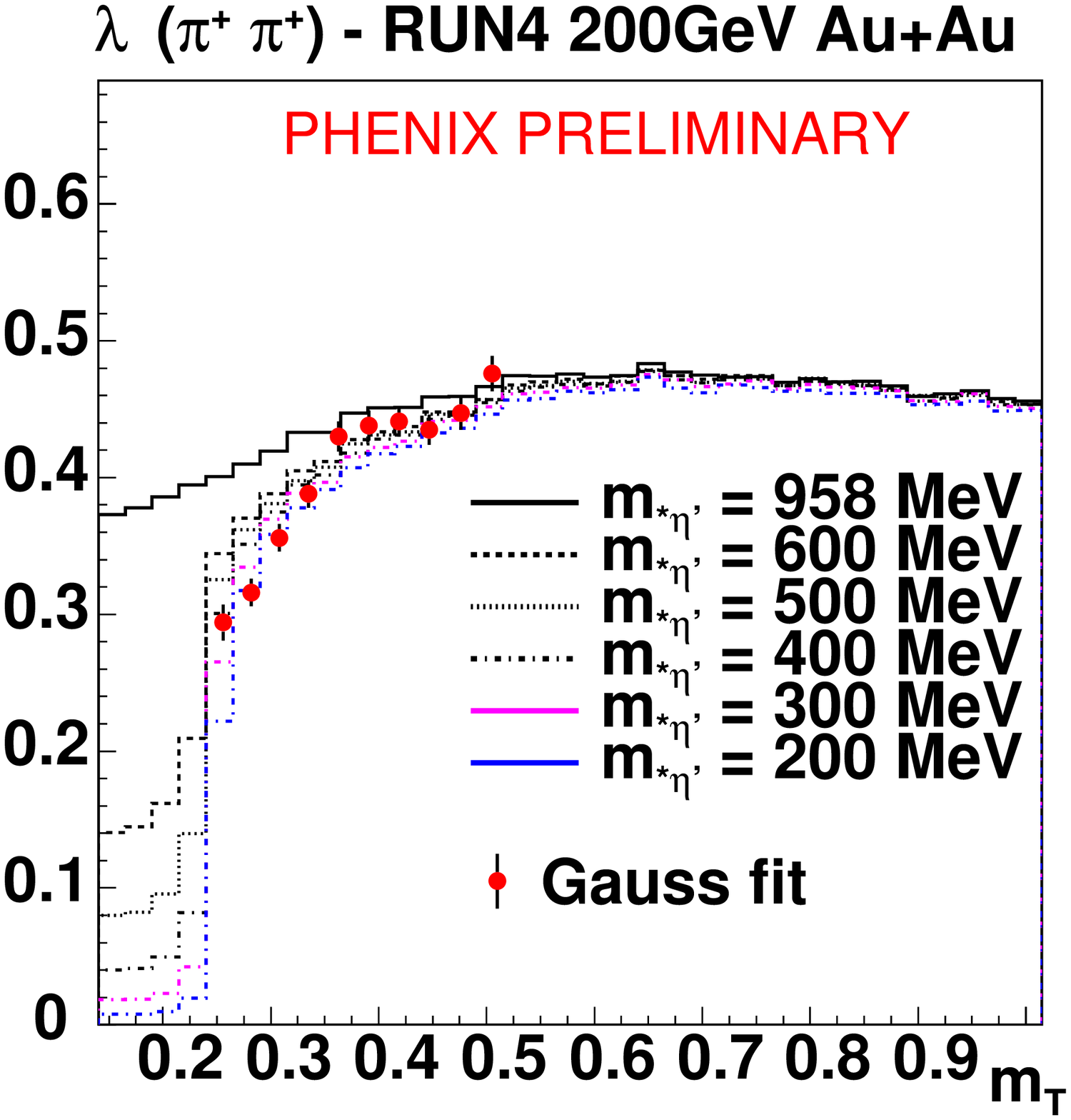}
   \includegraphics[width=1.9in]{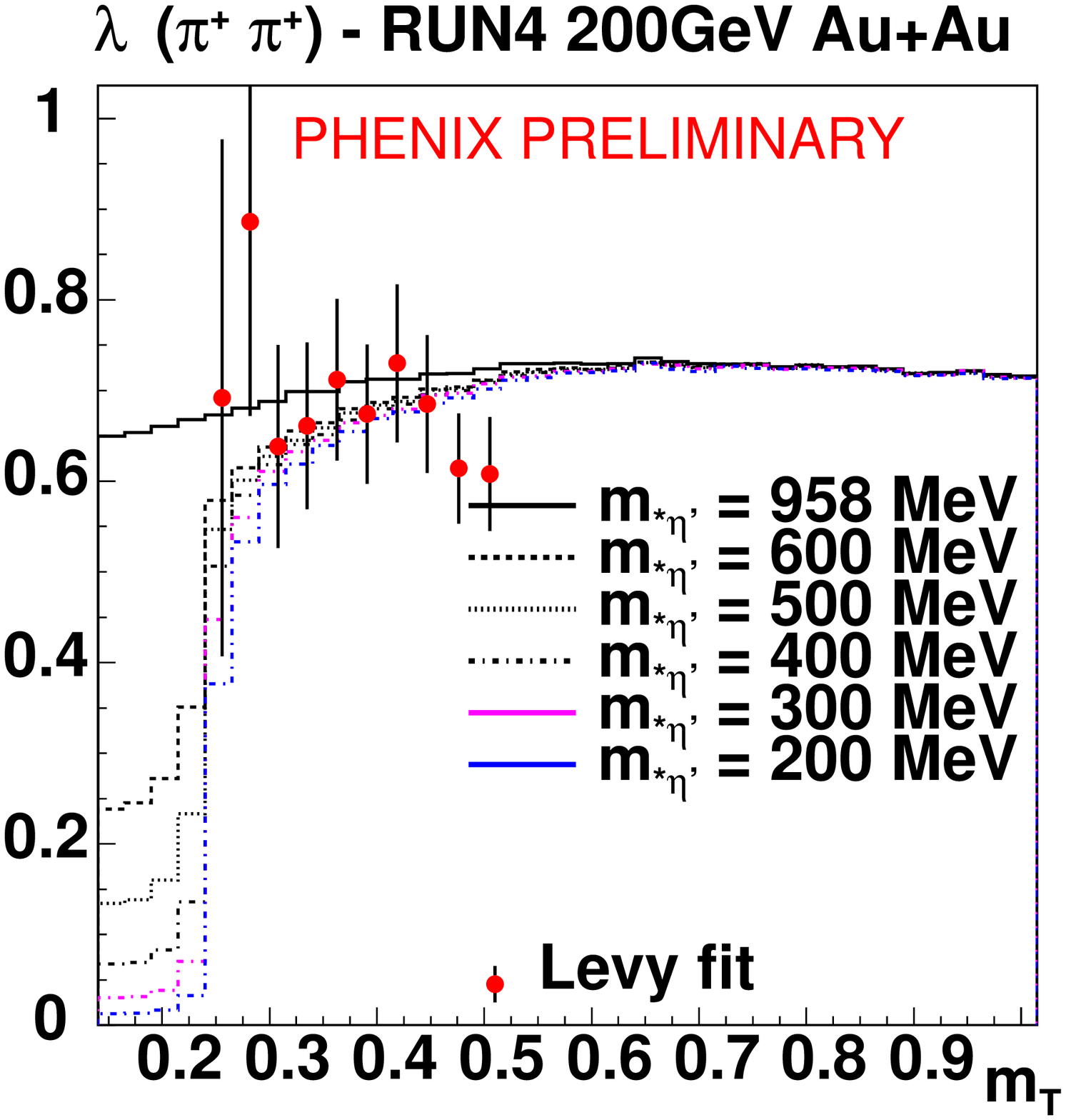}
   \includegraphics[width=1.9in]{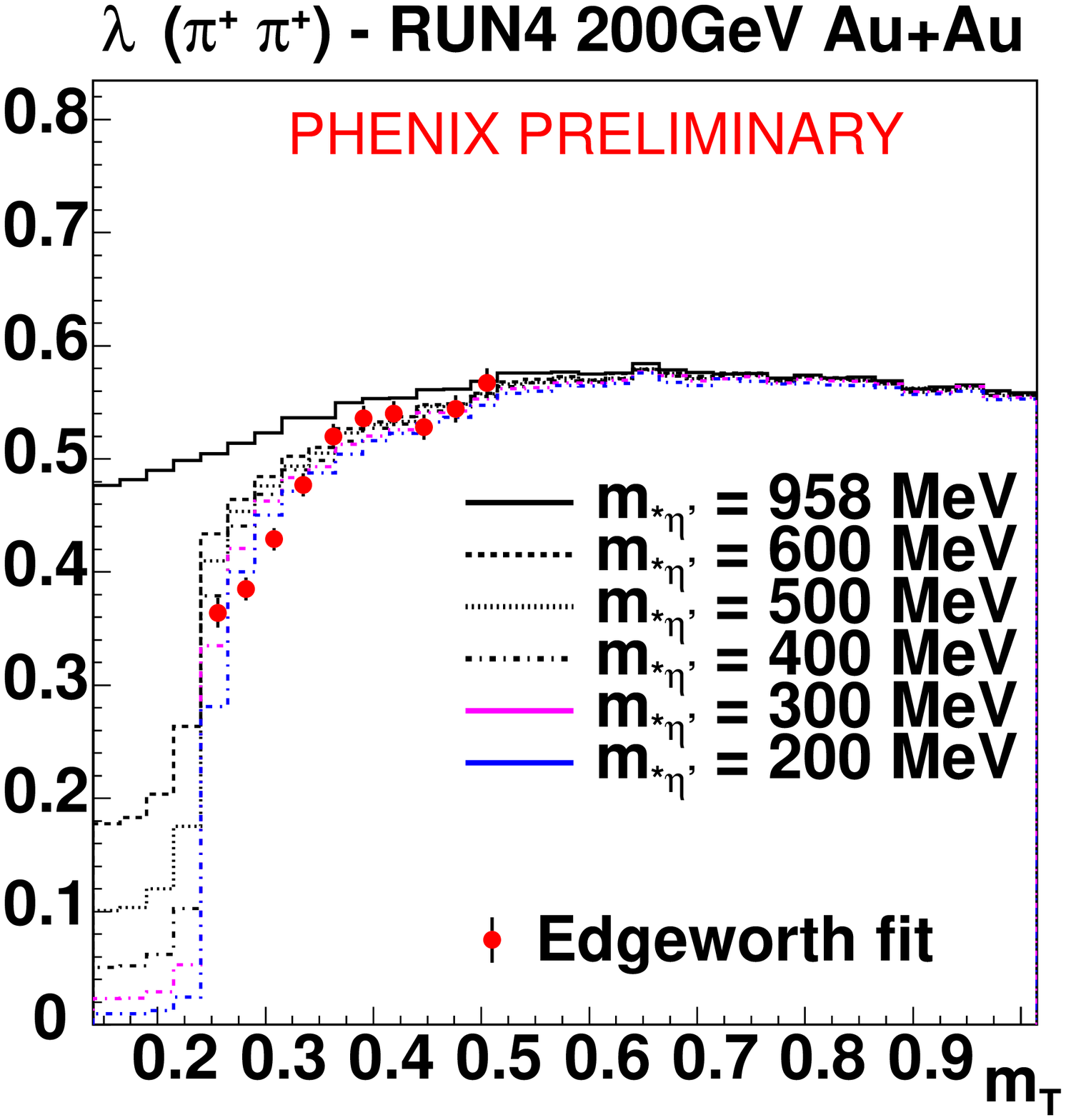}
   \caption{Measured $\lambda(m_t)$ compared to calculations using the model of ref.~\cite{Vance:1998wd}
            with various $\eta'$ mass values, and the Fritiof results for the composition
            of the long-lived resonances.}\label{f:ua1}
   \end{center}
\end{figure}

The $m_t$ dependence of $\lambda$ can be used to extract
information on the mass-reduction of the $\eta$' meson, a signal
of the U$_\textrm{A}(1)$ symmetry restoration in hot and dense
matter (see ref.~\cite{Vance:1998wd}). A comparison of the
measurements of fig.~\ref{f:shapeall} with model calculations of
ref.~\cite{Vance:1998wd} using Fritiof results for the composition
of the long-lived resonances and a variation of the $\eta'$ mass
is presented in fig.~\ref{f:ua1}.

\section{Conclusion}

From simultaneous analysis of two- and three-pion correlations we
obtained an allowed region of partial coherence versus core
fraction. Nonzero partial coherence is allowed when having larger
core size, at total incoherence ($p_c=0$) the core fraction is
slightly higher than previous analysis using Gaussian fits (see
refs.~\cite{Csorgo:1998tn,Boggild:1999tu,Biyajima:2003ey,Adler:2004rq}).
This is due to the fact that we used Levy fits, the only
investigated one that gave acceptable confidence levels.

We also analyzed the $m_t$ dependence of the $\lambda$ parameter
obtained from various fits. Gauss fit results agree with former
PHENIX measurements (see ref.~\cite{Adler:2004rq}). Regarding
U$_A$(1) symmetry restoration, we conclude that at present,
results are critically dependent on our understanding of
statistical and systematic errors, and additional analysis is
required to make a definitive statement.

\end{document}